\RequirePackage{fix-cm}

\documentclass[11pt,a4paper]{article}
\usepackage{bm}
\usepackage{latexsym}
\usepackage{dcolumn}
\usepackage{amsmath,amsfonts,amssymb}
\usepackage{graphicx,epsfig}
\usepackage{psfrag}
\usepackage{amsthm}
\usepackage{color}
\usepackage{cite}

\usepackage{nccmath}
\usepackage{moresize}
\usepackage{enumerate}
\usepackage[hang,flushmargin]{footmisc}
\usepackage[titletoc,toc]{appendix}
\usepackage{lipsum}
\usepackage{subcaption}
\usepackage{epstopdf} 
\usepackage{hyperref}
\usepackage{rotating}
\usepackage{multirow}
\usepackage{mathtools}

\usepackage{stackengine}
\usepackage{scalerel}

\newcommand{\be}{\begin{equation}}
\newcommand{\ee}{\end{equation}}
\newcommand{\bea}{\begin{eqnarray}}
\newcommand{\eea}{\end{eqnarray}}
\newcommand{\bse}{\begin{subequations}}
\newcommand{\ese}{\end{subequations}}
\newcommand{\bce}{\begin{center}}
\newcommand{\ece}{\end{center}}
\newcommand{\bfg}{\begin{figure}}
\newcommand{\efg}{\end{figure}}
\newcommand{\bit}{\begin{itemize}}
\newcommand{\eit}{\end{itemize}}
\newcommand{\bed}{\begin{description}}
\newcommand{\eed}{\end{description}}
\newcommand{\ben}{\begin{enumerate}}
\newcommand{\een}{\end{enumerate}}
\newcommand{\nn}{\nonumber}

\newcommand{\pa}{\partial}
\newcommand{\fr}{\frac}
\newcommand{\sq}{\sqrt}
\newcommand{\no}{\noindent}

\def\le {\left}
\def\ri {\right}
\def\a  {\alpha}
\def\b  {\beta}
\def\c  {\gamma}
\def\C  {\Gamma}
\def\d  {\delta}
\def\D  {\Delta}
\def\e  {\epsilon}

\def\k  {\kappa}

\def\L  {\Lambda}
\def\m  {\mu}
\def\n  {\nu}

\def\O  {\Omega}
\def\p  {\pi}

\def\r  {\rho}

\def\s  {\sigma}

\newcommand{\cC}{\mathcal C}

\newcommand{\cH}{\mathcal H}

\newcommand{\cL}{\mathcal L}

\newcommand{\cR}{\mathcal R}

\newcommand{\cV}{\mathcal V}

\newcommand{\Ct}{\widetilde{C}}

\newcommand{\ha}{\hat{\a}}
\newcommand{\hb}{\hat{\b}}
\newcommand{\hc}{\hat{\c}}
\newcommand{\hmu}{\hat{\mu}}
\newcommand{\hn}{\hat{\n}}
\newcommand{\hs}{\hat{\s}}

\newcommand{\vE}{\vec{\pmb E}}
\newcommand{\vB}{\vec{\pmb B}}
\newcommand{\vD}{\vec{\pmb D}}
\newcommand{\vH}{\vec{\pmb H}}

\newcommand{\Geff}{G_{\text{\scriptsize eff}}}

\newcommand{\bdm}{\begin{displaymath}}
\newcommand{\edm}{\end{displaymath}}

\DeclareMathSizes{12}{12}{5}{5}

\begin{document}

\markboth{Saurya Das and Sourav Sur}
{A Unified Cosmological Dark Sector from a Bose-Einstein Condensate}

\title{\LARGE{Varying Newton's constant: a cure for gravitational maladies?}}

\author{Saurya Das\footnote{Email: saurya.das@uleth.ca} \\ 
{\normalsize \em Theoretical Physics Group, Department of Physics and Astronomy,}\\
{\normalsize \em University of Lethbridge, 4401 University 
Drive, Lethbridge, Alberta T1K 3M4, Canada}\\ \\
Sourav Sur\footnote{Corresponding Author. Email: sourav@physics.du.ac.in} \\ 
{\normalsize \em Department of Physics and Astrophysics}\\
{\normalsize \em University of Delhi, Delhi - 110007, 
India}}

\date{}
\maketitle


\begin{abstract}
We show that a slowly varying Newton's constant, consistent with existing bounds, can potentially explain a host of 
%
observations pertaining to gravitational effects or phenomena across distances spanning from planetary to the cosmological,  
relying neither on the existence of Dark Matter or (and) Dark Energy, nor on any expected high proportions of either of them in the Universe. It may also have implications at very short distances or quantum gravity scales.
%
\end{abstract}

\vspace{10pt}
\no
{\it Keywords:} Varying Newton's constant, Modified Gravity, Gravitational Lensing, Dark Energy and Dark Matter, Cosmology of theories beyond the SM.




\section{Introduction}

Newton's theory of gravity and its subsequent relativistic formulation by Einstein, viz. General Relativity (GR), are among the most successful physical theories in explaining gravitational phenomena spanning from the planetary scale all the way to the galactic and cosmological scales.
However, despite being largely successful, these theories have significant limitations, notably the following:
\bed 
\item (i) the need for an as-yet unidentified entity known as Dark Matter (DM) for explaining gravitational phenomena observed at the galactic scales and beyond, 
\item (ii) the requirement of a small, positive cosmological constant or 
%
more generally, Dark Energy (DE), also of an unknown origin, in order to explain observations at the cosmological scales, and 
\item (iii) the challenge in constructing a self-consistent quantum theory of gravity, 
notwithstanding the success of perturbative quantum field theory in describing the other fundamental forces of nature.
\eed 
%

Our objective in this paper is to explore potential resolutions to some of these issues (specifically the first two, which would be relevant for the third one as well) by making a reasonable proposal that the Newton's constant $G$ may not strictly be a `constant', but may instead vary slowly with space and possibly with time as well. 
Of course, any such proposal must need to be consistent with the known experimental bounds of observable effects ranging from the planetary and astrophysical scales to the cosmological scale. 

While there exist huge volumes of literature on gravitational theories incorporating a variation or a scale-dependence of $G$, which we shall discuss in some detail in the section \ref{sec:VarG} below, the one we propose is a simple Taylor expansion in the radial coordinate $r$, about an arbitrarily chosen origin $r = 0$. Presuming the zeroth order term in the expansion to be the experimentally measured Newton's constant at terrestrial and planetary scales, the physical relevance of a few higher order terms can be realized at once, under certain stipulations. For instance, the first order term, if stipulated to be positive definite, modifies the Newtonian gravitational potential by a logarithmic term, 
well-known for providing an alternative to the dark matter (DM). 
The second order term, on the other hand, modifies the Newtonian force by a constant term, which can be interpreted as that due to a gravitational retardation, similar to what one encounters in electrodynamics.
The third order term in the expansion, if stipulated to be negative definite, leads to the perception of an effective positive cosmological constant, the most favoured candidate for dark energy (DE). 
Beyond the third order though, any of the individual expansion terms in $G(r)$ cannot be readily interpreted, regardless of any stipulation. Henceforth, limiting our attention to this order in the expansion, we endeavor to examine whether such an expansion can lead to any observable effect(s) of significance in astrophysics and cosmology.  

The paper is organized as follows: starting with the general layout of the gravitational theory involving the radial expansion of $G$ in section \ref{sec:VarG}, we discuss the consequences at different length scales, specifically, in accounting for astrophysical effects, such as the flatness of the rotation curves of galaxies, the galactic or cluster mass estimation via gravitational lensing and from the perspective of the virial theorem, as well as the cosmological dynamics, in section \ref{sec:VarG-effects}. A detailed cosmological study is then carried out in this context, albeit in a quasi-Newtonian framework, in section \ref{sec:VarG-cosm}, and consequently the parameter estimation using standard statistical techniques and the combination of two widely used observational data-sets, for various emerging cosmological scenarios. A comparative study of such scenarios is also done by resorting to three statistical indicators, viz. the minimized $\chi^2$ per degree of freedom (dof), AIC and BIC. Possible small scale effects of the varying $G$ are then discussed briefly in section \ref{sec:VarGshortdist}, before summarizing and concluding with a discussion on the open issues, work that remains to be done and works that are in progress, in section \ref{sec:Concl}. A possible embedding of the varying $G$ in a covariant framework is illustrated in the Appendix.

\section{Varying $G$ and corrections to the Newtonian Gravitational Potential} \label{sec:VarG}
Considering, for brevity, spatial
variations that are spherically symmetric, we can write the following without loss of generality:
%
\be \label{g1}
G (r) =\, G_0 +\, G_1\, r +\, G_2\,r^2 +\, G_3\, r^3 + \dots \,, 
\ee
%
where the coefficients $\,G_0,\, G_1,\, G_2, \dots$ can be positive or negative, depending on the physical situation. In fact, the expansion can be made around an arbitrarily chosen origin $r=0$, 
%
%
with $G_0$ considered to be positive definite, being the `observed' value of the Newton's constant at terrestrial and planetary scales. The other coefficients may conveniently be expressed in terms of $G_0$ as
\be \label{Gn}
|G_n| \equiv \frac{G_0}{(r_n)^n}~, \quad (n = 1,2,\dots) \,,
\ee
where $r_n$ denotes the length scale at which the respective terms in Eq.(\ref{g1}) become dominant. For example, $r_1$ can be taken to be the galactic scale, $r_2$ an intermediate scale and $r_3$ the cosmological scale, or more specifically, $r_1 \simeq 10^{21}\,$m and $r_3 \lesssim L_0$, with $L_0 \approx 10^{26}\,$m being the Hubble radius, and $r_1 < r_2 < r_3$.

While, admittedly, there is the requirement of an unambiguous assertion of the signs of the coefficients $G_n$, and also of the explanation of the expansion (\ref{g1}) altogether (despite there being certain concrete theoretical formalisms in existence, which we shall mention below), a reasonable physical picture can be reckoned if $G_1$ is positive definite, so that gravity at the galactic scale remains attractive (i.e. galaxies can form), and $G_3$ is negative definite, in order that an effective positive cosmological constant is perceived, thus leading to an accelerated phase of expansion of the Universe at late times. 
%
The signs of the other coefficients in Eq.(\ref{g1}), viz. $G_2, G_4, G_5, \dots$, may be positive or negative, and we do not make any specific choice for any of them beforehand --- their assertion is to be left entirely to the statistical parametric fits with observational data in specific contexts, such the late time cosmological background evolution, to be studied in this paper.   
%
%
%
%

Let us also remind that one always has a substantial room for making an expansion such as that in Eq.\,(\ref{g1}), as long as
\be 
\frac{|G(r) - G_0|}{G_0} \ll 10^{-N} \,, \quad \forall ~r < L_0 \,, 
\label{bound}
\ee
where $N \gtrsim 5$ 
\cite{mikkelsen}.
%
This however does not necessarily imply that an individual correction term in Eq.\,(\ref{g1}) would always have a subdued effect in astrophysical phenomena or the cosmological evolution. The effect may be significant at a particular length scale, $r_n$, since the suppression factor $(r_n)^{-n}$ in the respective $G_n$ may not appear on the whole in a given computation.

Substituting Eq.\,(\ref{g1}) in the Newtonian force per unit mass, due to a gravitating object of mass $M$ at a distance $r$, we get
%
\bea \label{er}
E (r) &=&  -\, \frac{GM}{r^2}  
\nn \\
%
&=&
-\, \fr{G_0 M}{r^2} -\, \fr{G_1 M} r -\, G_2 M -\, G_3 M \,r - \dots \,, 
\eea
%
%
from which it is easy to see that a particular 
$G_n M \,r^{n-2}$ term dominates in the range $r_n < r < r_{n+1}$, 
since by Eq.\,(\ref{Gn}), the magnitudes of the coefficients $G_n$ decrease rapidly with $n$.
%
One obtains the corresponding gravitational potential as
\bea \label{pot}
V (r) &=& -\, \int d \vec r \cdot \vec E (r) \nn \\
&=& 
- \fr{G_0 M} r \le[1 -\, \fr r {r_1} \ln \fr r {r_\ell} \pm\, \fr{r^2}{(r_2)^2} +\, \fr{r^3}{2 (r_3)^3} \pm\, \dots\ri] \,, 
\eea
%
where $r_\ell$ is a constant --- an arbitrary scale of the logarithm, and we have taken $G_3 = - G_0/(r_3)^3$ for the reason mentioned above. 

Expansions (\ref{er}) and (\ref{pot}) are consistent with the bound (\ref{bound}). The logarithmic correction term in the potential (\ref{pot}) is well-known and has been considered by numerous authors, see e.g. ref. 
\cite{sivaram}.
%
%
Generic variations of the Newton's constant $G$ have also been proposed in various covariant theoretical formulations of gravity, starting with the Brans-Dicke theory to the plethora of its generalizations, commonly classified as the scalar-tensor theories 
\cite{bd1,bd2,FM-ST,frni-ST}. 
In fact, almost all of the so-called {\em modified gravity} theories, e.g. $f(R)$ and its variants 
\cite{fr,CFPS-mg,papa-ed,NOO-mg}, 
have scalar-tensor equivalents. Even rigorous formulations, such as the tensor-vector-scalar (TeVeS) theory 
\cite{bek-TeVeS,MSY-TeVeS,BS-TeVeS}, 
turned out to be suitable for a covariant generalization of the modified Newtonian dynamics (MOND) 
\cite{mil-MOND,MS-MOND,bek-MOND},
well-known for providing an alternative to cold dark matter (CDM). On the other hand, non-local gravitational theories in particular, such as that in refs. 
\cite{HM-NLG,BCHM-NLG,mash-NLG}, 
provide concrete covariant formulations for a $G$-variation precisely in the form of the logarithmic potential term shown in Eq.\,(\ref{pot}). Among recently proposed formulations of similar sort, it is worth mentioning ref. 
\cite{nash} 
wherein a gravitational self energy term is argued from certain standpoints.

Quite intriguing, in this context, are quantum field theoretic approaches, notably, the ones dealing with the renormalization group (RG) flow in gravitational formulations
\cite{RW-QGRG,SSS-QGRG,GNS1-QGRG,GNS2-QGRG,RLS-QGRG}.
These essentially amount to propositions for effective quantum gravity theories wherein the dependence of $G$ on the RG group scale parameter $\m$ is of importance. The quantum correction to the effective gravitational potential at low energies, thus obtained, is similar to the logarithmic term in Eq.\,(\ref{pot}) (see also the works on the quantum corrections to the Newtonian potential from the point of view of the gauge fixing independence
\cite{NMS-QGGF,DM-QGGF}).
While the astrophysical consequences of such a scale dependence of $G$, particularly the flatness of the galaxy rotation curves, have been explored in detail in refs.
\cite{RW-QGRG,RLS-QGRG},
intensive studies of the cosmological implications of the same (alongwith that of a varying cosmological constant $\L$) have been carried out in refs.
\cite{SSS-QGRG,BHMR-RGcosm,SS-RGcosm,AKLR-RGcosm}
and with more generality and refinements in the recent paper
\cite{BRS-RGcosm}
(see also the references therein, for further details).

\section{Consequences of the varying $G$ at different scales} \label{sec:VarG-effects}

For the accountability or the degree of robustness of the proposed radial variation of $G$ in Eq.\,(\ref{g1}),
let us pinpoint the supporting arguments or evidences from the perspective of the astrophysical observations as well as the emerging cosmological picture(s), as follows: 
%
\ben[(a)]
\item Consider first, the galaxy {\bf rotation curves}, in the range $r_1 \leq r \leq r_2$. By equating the gravitational force on a unit mass, Eq.\,(\ref{er}), to its centrifugal acceleration $v^2/r$, one obtains
%
%
\be
%
v =\, \sq{G_0 M \le[\fr 1 r +\, \fr 1 {r_1} \mp\, \fr r {(r_2)^2} - \fr{r^2}{(r_3)^3} \mp\, \dots\ri]} \,.
\label{velo1}
\ee
%
Within the aforementioned range, the second term on the right hand side
dominates, leading to
\be
v \approx \sqrt{\frac{G_0 M}{r_1}} = \mbox{constant}.
\ee 
In fact, setting $M \simeq 10^{42}\,$Kg, a typical galaxy mass, 
one gets $v \simeq 10^5\,$m/s, which is indeed the typical velocity of the rotation curves. 
\item Consider next, the {\bf gravitational lensing}, also in the above range.
As shown in ref.
\cite{dsphysopen}, 
%
the deflection angle of light rays and the Einstein radius for a gravitational lens of mass $M$ comparable to that of a galaxy or cluster, are respectively given by  
\bea 
\delta 
%
&\equiv& \fr{4 G_0 M'}{b \, c^2} \,, 
\label{delta4} \\
\theta_E 
%
&=& \fr 2 c \, \sq{\le(\fr{D_{LS}}{D_L D_S}\ri) G_0 M'} \,,
\label{einstein1}
\eea
where $b$ denotes the impact parameter, $D_S,D_L$ and $D_{LS}$ are the distances from the observer to the source and the lens and the distance from the source to the lens respectively, and
%
\be 
M' \equiv \le( 1 + \e\ri) M \,,
\ee 
is the apparent mass, with 
%
\be \label{epsilon}
\e =\, \fr{\pi b}{2 r_1} \simeq\, 1~\mbox{to}~10\,
\ee 
at these distances.
%
In other words, the apparent mass $M'$ estimated via lensing, 
if considered without the $1/r$ term in Eq.(\ref{er}) or the logarithmic term in Eq.(\ref{pot}), can exceed the actual gravitating mass $M$
by a factor of up to $10$. 
%
This result applies, in general, to all galaxies and clusters, including the 
bullet cluster
%
\cite{bullet1,bullet2}.
%
We will comment more on it in the concluding part of this paper.
\item 
Subsequently consider, the {\bf virial theorem}, which is a fundamental tool used to determine galaxy masses based on average velocity data. It is not difficult to show that the inclusion of the additional terms in Eqs.(\ref{er}) and (\ref{pot}) modifies the theorem \cite{binney}, and one now has:
\be
\langle v^2 \rangle = G_0 M \left( \frac{1}{b} + \frac{1}{r_1}\right) \simeq
\frac{G_0M'}{b}~,
\ee
where $b$ represents a typical galaxy size.
%
As in the case (b) above, for gravitational lensing, a mass over-estimation, of up to $10$ times, may result upon neglecting such additional terms, which again shows the importance of the latter when calculating galaxy or cluster masses.
%
\item  
Consider now, the term containing $r_2$ in Eqs.\,(\ref{er}) and (\ref{pot}) in the range $r_2 < r < r_3$. This corresponds to an additional constant force, which may be identified with a force of {\em retardation}, as a result of the finite speed of propagation of gravitational signals, as shown in ref.
\cite{yahalom}. 
While clearly more work needs to be done to fully comprehend the implication of this force, here we simply point out that such a force predicted by a varying Newton's constant is supported by independent considerations. 
\item Finally, consider the range $r_3 \leq r \leq L_0$, where $L_0 = H_0^{-1}$ is the Hubble radius, with $H_0$ being the Hubble's constant. This range typifies the {\em cosmological} length scale, at which the $G_3$ term in Eq.\,(\ref{er}), or correspondingly, the term containing $r_3$ in Eq.\,(\ref{pot}), becomes dominant. With $G_3$ taken to be negative, such a term can be identified as an effective positive cosmological constant $\L$, up to a numerical factor.
%
As such, one can realize a viable cosmological scenario exhibiting a late-time accelerating regime, albeit with the intriguing aspect of having the other terms in the varying $G$, notably the ones leading to the logarithmic term and the constant force term in Eq.\,(\ref{pot}), playing roles of significance towards complementing not only the usual pressure-less ({\em dust}-like) cosmological matter, but also the effective dark energy, making the latter {\em dynamical}. 
While a full-fledged analysis in this respect obviously
requires an underlying covariant formulation incorporating the $G$-variation of the sort we propose in Eq.\,(\ref{g1}), let us point out that an inherently non-relativistic quasi-Newtonian approach may serve the purpose, at least technically, for the background level cosmological evolution. In other words, one may bypass the ambiguity in deciding which of such covariant formulations to resort to, by working in a quasi-Newtonian cosmological framework, which despite being somewhat heuristic, leads in general to the same background evolution equations, viz. the Friedmann equations, as in relativistic cosmology\footnote{At the level of the cosmological density perturbations though, the quasi-Newtonian formalism is expected to differ significantly from the relativistic one. Nevertheless, while such a perturbative study, in an appropriate covariant formulation for the $G$-variation, is being carried out in some ongoing works
\cite{ds2,SAD},
we keep it beyond the scope of this paper.}. Limiting ourselves to the study of the background level cosmology, we therefore take the quasi-Newtonian approach to work out explicitly, in the following section, how the individual terms in the varying $G$ contribute to the effective Friedmann equations. As we shall see, the cosmological parametric estimation, consequently carried out therein, using the data available from the type-Ia Supernovae observations and the measurements of the Hubble parameter using cosmic chronometry, distance ladder, etc. indicate a significant reduction of the amount of the cold dark matter (CDM) content of the Universe, compared to the concordant $\L$CDM model.   
%
\een

\section{Cosmological implications of the varying $G$ and parametric estimates} \label{sec:VarG-cosm}

Following refs.
\cite{inverno,liddle,rai,weinberg}, 
let us consider the Universe as a spherically symmetric distribution of galaxies.
By the cosmological principle, the galactic motion is purely radial, and the radius vector of a test galaxy with respect to a chosen (fixed) origin can always be written as
\be \label{radvec}
\vec{\pmb r}(t) =\, a(t) \, \ell(t_0) \, \widehat{\pmb r} \,,
\ee 
where $t_0$ is some reference epoch which we may take, for convenience, as the present epoch, $\ell(t_0)$ is the characteristic length scale, and $a(t)$ is a universal function of time, namely, the `scale factor'. 
So, the kinetic energy of the test galaxy, per unit mass, can be expressed as
\be \label{KE}
E_k =\, \fr 1 2 \, \dot{\vec{\pmb r}}^2 =\, \fr{r_0^2} 2 \, \fr{\dot a^2}{a_0^2} \,,
\ee 
where the dot $(\cdot) \equiv d/dt$, and we have denoted: $a_0 \equiv a(t_0) = r_0/\ell_0\,$, $\, r_0 \equiv r(t_0)$, $\ell_0 \equiv \ell(t_0)$. 
%

On the other hand, keeping for brevity, terms up to order $r^3$ in the expansion for $G(r)$, given by Eq.\,(\ref{g1}),  we have the Eq.\,(\ref{pot}) for the gravitational potential reducing to 
\be \label{PE}
V =\, -\, \fr{4 \pi G_0 r_0^2 a^2}{3 a_0^2} \le[1 -\, \a  \ln \fr a {a_\ell} +\, \b a^2 +\, \c a^3\ri] \rho \,, \quad
\ee 
where
\be 
\a \equiv\, \fr{a_0 r_0}{r_1}\,, \quad \b \equiv \pm \le(\fr{a_0 r_0}{r_2}\ri)^2 \,, \quad \c \equiv\, \fr 1 2 \le(\fr{a_0 r_0}{r_3}\ri)^3 \,, \quad a_\ell \equiv\, \fr{a_0 r_\ell}{r_0} \,, 
\ee 
and $\rho(t) = M/\cV(t) \,$ is the mass density of the distribution, within the spherical volume $\cV(t) = 4 \pi \ell_0^3 a^3(t)/3$.

The total energy of the system, per unit mass of the test galaxy, is expressed as
\be \label{TotE}
E 
=\, - \fr 1 2 \,k c^2 \,,
\ee 
where $k$ is a constant, which can either be $0$ (when $E = 0$) or $\pm 1$ (under a suitable scaling, for $E \lessgtr 0$), and is often regarded as the Newtonian analogue of the spatial curvature constant, at least mathematically.

Setting $E = E_k + V\,$, and defining $H = \dot a/a$, the analogue of the Hubble parameter, we get
\be \label{hub1}
H^2 + \fr k {a^2} =\, \fr{8 \pi G_0} 3 \le[1  - \a\, a \ln \fr a {a_\ell} + \b a^2 + \c a^3\ri] \r \,,
\ee 
upon a rescaling of $k$, by a dimensionless factor $(a_0 c/r_0)^2\,$,
which however does not forbid $k$ to be suitably scaled further to $\pm 1$ for $E \lessgtr 0$. Thus, one may treat $k$ as still the spatial curvature constant analogue, and consequently, Eq.\,(\ref{hub1}) as the analogue of the Friedmann equation, modified by the spatial variation of $G$.

Let us now resort to the general dilution of the matter density $\r$ with the scale factor $a$. A simple way of seeing this is by considering the first law of thermodynamics for an adiabatic expansion for a perfect fluid of energy $U = \rho \cV$ and pressure $p$
\cite{rai}:
\be \label{td1}
dU +\, p \, d\cV =\, 0 \,,
\ee 
which can be re-expressed as
\be \label{consv}
\dot \rho +\, 3 H \le(\rho + p\ri) =\, 0 \,,
\ee 
since $\cV \propto a^3$. 
%
Note that the consideration of Eq.\,(\ref{td1}) is in realm of a {\em quasi-Newtonian} formalism, rather than a purely Newtonian one which, although has provision, does not incorporate Eq.\,(\ref{td1}) by default.  

Differentiating Eq.\,(\ref{hub1}) with respect to $t$ and using Eq.\,(\ref{consv}), we get
%
%
%
\bea \label{R-eq}
\fr{\ddot a} a &=& \dot H +\, H^2 \nn \\ 
&=& -\, \fr{4 \pi G_0} 3 \bigg[\rho + 3p + \le(\rho - 3p \, \ln \fr a {a_\ell}\ri) \a a - \le(\rho - 3p\ri) \b a^2 +\, 3 \c a^3 p \bigg] \,, 
\eea 
which can be regarded as the analogue of the standard cosmological version of the Raychaudhuri equation, modified by the $G$-variation (up to the third order).

Considering further, the matter content in the distribution to be purely non-relativistic, i.e. with negligible pressure, Eq.\,(\ref{consv}) can be solved to give the matter density as 
\be \label{rho_m}
\rho (t) =\, \rho_0 \, a^{-3} (t) \,,
\ee 
with the value $\rho_0$ at the present epoch $(t = t_0)$ at which we set $a = 1$ without loss of generality.
%
Consequently, the last term in Eq.\,(\ref{hub1}) becomes a constant,  which (modulo a numerical factor of $3$) can be identified as an effective cosmological constant 
\be \label{CC}
\L \equiv \, 8 \p G_0 \c \r_0 \,. 
\ee 
Finally, considering the following density parameters (or specifically, their values at the present epoch $t_0$): 
\bea \label{Omega-def}
&&\O^{(m)}_0 =\, \fr{8 \pi G_0 \rho_0}{3 H_0^2} \,, ~~ \O^{(\L)}_0 =\, \fr \L {3 H_0^2} \,, ~~ \O^{(k)}_0 =\, -\, \fr k {H_0^2} \,, \nn\\
&& \O^{(\a)}_0 =\, \fr{8 \pi G_0 \a \rho_0}{3 H_0^2} \,, ~~ \O^{(\b)}_0 =\, - \fr{8 \pi G_0 \b \rho_0}{3 H_0^2} \,, ~~ 
\eea 
and using Eq.\,(\ref{rho_m})
%
%
we re-write Eq.\,(\ref{hub1}) as
\be \label{hub1a}
\le[\fr{H(a)}{H_0}\ri]^2 =\, \fr{\O^{(m)}_0}{a^3} + \fr{\O^{(k)}_0}{a^2} -\, \fr{\O^{(\a)}_0}{a^2} \ln \fr a {a_\ell} +\, \fr{\O^{(\b)}_0} a +\, \O^{(\L)}_0 \,,
\ee
where $H_0 \equiv H(t_0)\,$ denotes the Hubble's constant.

Note that the third term in the above equation (\ref{hub1a}) can be split into two terms --- one varying as $(\ln a)/a^2$ and the other varying as $1/a^2$. The latter actually augments the second term in Eq.\,(\ref{hub1a}) to give a term corresponding to an effective spatial curvature constant analogue, say $k'$, the value of whose density parameter at the present epoch $t_0$ is
\be \label{Omega_k'}
\O^{(k')}_0 :=\, \O^{(k)}_0 +\, \O^{(\a)}_0 \, \ln a_\ell \,.
\ee 
Consequently, Eq.\,(\ref{hub1a}) takes the form
\bea \label{hub1b}
\le[\fr{H(z)}{H_0}\ri]^2
&=& \O_0^{(m)}\,(1+z)^3 +\, \O_0^{(k')}\,(1+z)^2 \nn \\
&+& \O_0^{(\a)}\,(1+z)^2 \, \ln(1+z) +\, \O_0^{(\b)}\,(1+z) +\, \O_0^{(\L)} \,,
\eea 
when expressed in terms of the redshift $z = a^{-1} - 1$.

Eliminating $\O_0^{(\L)}$ via the condition one gets from Eq.\,(\ref{hub1b}) at $z = 0$, viz.
\be \label{constraint}
\O_0^{(m)} +\, \O_0^{(k')} +\, \O_0^{(\b)} +\, \O_0^{(\L)} =\, 1 \,,
\ee 
a maximum likelihood analysis can be carried out for the other density parameters ($\O_0^{(m)}$, $\O_0^{(k')}$, $\O_0^{(\a)}$ and $\O_0^{(\b)}$), as well as for the Hubble's constant $H_0$, using observational data. We do so by adopting the well-known Metropolis-Hastings algorithm for the Markov-chain Monte Carlo (MCMC) random probabilistic exploration, with a combination of two widely used data-sets, viz. the Pantheon+SH0ES data-set consisting of $1701$ light curves of Supernovae type Ia (SNeIa) 
\cite{scol-Pan+,brout-Pan+,riess-SH0ES,Pan+}, 
and the observational Hubble data-set (OHD), the full compilation of which presently consists of $51$ measurements of $H(z)$, $34$ of them using cosmic chronometry (CC) 
\cite{mors-OHD,MPCJM-OHD,BMC-OHD,tomasetti-OHD},
$3$ from SNeIa distance ladder
\cite{riess-SH0ES,riess-Lad_OHD},
$7$ from radial Baryon Acoustic Oscillation (BAO) signal size in galaxy distribution
\cite{GCH-BAO-Gal_OHD,blake-BAO-Gal_OHD},
and the remaining $7$ ($4$ and $3$ respectively) from Baryon Oscillation Spectroscopic Survey (eBOSS)
\cite{zhao-BAO-QSO_OHD}
and Lyman-$\a$ forest emission in quasar spectra
\cite{font-BAO-Lya_OHD,delubac-BAO-Lya_OHD,dong-BAO-Lya_OHD}
(see the complete OHD list, and the duly cited references, in the Table 1 of the Appendix C of the recent paper
\cite{monjo}).
%

The likelihood function is in general given by $\, \cL = \exp\le(- \chi^2/2\ri)$, where 
\be \label{chisq}
\chi^2 =\, \sum_{i,j=1}^N \D V_i \cdot C_{ij}^{-1} \cdot \D V_j \,,
\ee 
with $N$ being the size of a given sample, $\, \D V_i = V_{\text{\scriptsize obs}} (z_i) - V_{\text{\scriptsize th}} (z_i)\,$ the difference between the observed and the theoretically predicted values of the observable $V$ under consideration, at a particular measurement point $z_i\,$, and $C_{ij}$ denoting the elements of the corresponding covariance matrix. In the present analysis, $z_i\,$ denotes the redshift at which an observation is made, and the total $\chi^2$ is the sum of those for the two data-sets, viz. the Pantheon+SH0ES data-set and the OHD. While the observable $V$ for the former is the distance modulus $\m(z)$ of the Supernovae, it is nothing but the Hubble parameter $H(z)$ itself for the latter. 

%
%
\begin{table*}[t]
\caption{\footnotesize{Best fit values and $1\s$ confidence limits of the parameters of relevance, such as $H_0$, $\O_0^{(m)}$, $\O_0^{(k')}$, $\O_0^{(\a)}$ or(and) $\O_0^{(\b)}$, for various cases emerging from the varying $G$ scenario, using the Pantheon+SH0ES data combined with OHD. The minimized $\chi^2$ per degree of freedom (dof), as well as the AIC and BIC values, for the respective analyses are also quoted.}}
\begin{center}
\renewcommand{\arraystretch}{1.35}
\setlength{\tabcolsep}{3pt}
{
\begin{tabular}{|c c|c||c|c|c|c|c||c|c|c|}
\hline
\multicolumn{3}{|c||}{Specific} & \multicolumn{5}{c||}{Parametric estimates (Best fit \& $1\s$ limits)} & \multicolumn{3}{c|}{Statistics} \\
%
\cline{4-11}
%
\multicolumn{3}{|c||}{Scenarios} & $H_0$ & $\O_0^{(m)}$ & $\O_0^{(k')}$ & $\O_0^{(\a)}$ & $\O_0^{(\b)}$ & ${\chi^2_{\text{\scriptsize min}}}/{\text{\small dof}}$ &  AIC & BIC \\
\hline\hline
\multirow{3}{*}{\rotatebox[origin=c]{90}{Flat}} & \multirow{3}{*}{\rotatebox[origin=c]{90}{Universe}} & $F_0$ & $73.907_{-0.161}^{+0.159}$ & $0.254_{-0.008}^{+0.008}$ & --- & --- & --- & $1.046$ & $1838.799$ & $1849.736$ \\
\cline{3-11}
& & $F_1$ & $73.942_{-0.164}^{+0.165}$ & $0.167_{-0.094}^{+0.062}$ & --- & $0.230_{-0.163}^{+0.248}$ & --- & $1.045$ & $1840.807$ & $1857.212$ \\
\cline{3-11}
& & $F_2$ & $73.653_{-0.168}^{+0.162}$ & $0.147_{-0.087}^{+0.062}$ & --- & $0.229_{-0.164}^{+0.231}$ & $0.184_{-0.025}^{+0.012}$ & $1.034$ & $1823.318$ & $1845.192$ \\
\hline\hline
\multirow{3}{*}{\rotatebox[origin=c]{90}{Curved}} & \multirow{3}{*}{\rotatebox[origin=c]{90}{Universe}} & $C_0$ & $73.859_{-0.171}^{+0.163}$ & $0.245_{-0.010}^{+0.011}$ & $0.032_{-0.025}^{+0.013}$ & --- & --- & $1.043$ & $1837.020$ & $1853.425$ \\
\cline{3-11}
& & $C_1$ & $73.799_{-0.169}^{+0.165}$ & $0.145_{-0.084}^{+0.062}$ & $0.077_{-0.032}^{+0.017}$ & $0.224_{-0.161}^{+0.225}$ & --- & $1.041$ & $1836.281$ & $1858.155$ \\
\cline{3-11}
& & $C_2$ & $73.552_{-0.187}^{+0.183}$ & $0.137_{-0.080}^{+0.057}$ & $0.059_{-0.049}^{+0.028}$ & $0.212_{-0.149}^{+0.217}$ & $0.181_{-0.028}^{+0.015}$ & $1.032$ & $1823.052$ & $1850.394$ \\
\hline\hline
\end{tabular}
}
\end{center}
\label{LP-EstTable-2f} 
\end{table*} 
%
Table\,\ref{LP-EstTable-2f} quotes the best fit values and the corresponding $1\s$ margins of the parameters of relevance in the following scenarios arising out of varying $G$ formalism:
\bed 
\item $F_0\,$: with the pre-assignment $\, \O_0^{(k')} = 0 = \O_0^{(\a)} = \O_0^{(\b)}$, so that the relevant parameters are $\,H_0$ and $\O_0^{(m)}$, alongwith $\O_0^{(\L)} = 1 - \O_0^{(m)}$ (by Eq.\,(\ref{constraint})).
\item $F_1\,$: with the pre-assignment $\, \O_0^{(k')} = 0 = \O_0^{(\b)}$, so that the relevant parameters are $\,H_0, \O_0^{(m)}$ and $\O_0^{(\a)}$, alongwith $\O_0^{(\L)} = 1 - \O_0^{(m)}$ (by Eq.\,(\ref{constraint})).
\item $F_2\,$: with the pre-assignment $\, \O_0^{(k')} = 0$ only, so that the relevant parameters are $\,H_0, \O_0^{(m)}, \O_0^{(\a)}$ and $\O_0^{(\b)}$, alongwith $\O_0^{(\L)} = 1 - \O_0^{(m)} - \O_0^{(\b)}$ (by Eq.\,(\ref{constraint})).
\item $C_0\,$: with the pre-assignment $\, \O_0^{(a)} = 0 = \O_0^{(\b)}$, so that the relevant parameters are $\,H_0, \O_0^{(m)}$ and $\O_0^{(k')}$, alongwith $\O_0^{(\L)} = 1 - \O_0^{(m)} - \O_0^{(k')}$ (by Eq.\,(\ref{constraint})).
\item $C_1\,$: with the pre-assignment $\, \O_0^{(\b)} = 0$ only, so that the relevant parameters are $\,H_0, \O_0^{(m)}, \O_0^{(k')}$ and $\O_0^{(\a)}$ alongwith $\O_0^{(\L)} = 1 - \O_0^{(m)} - \O_0^{(k')}$ (by Eq.\,(\ref{constraint})).
\item $C_2\,$: with no pre-assignment, so that all the parameters $\,H_0, \O_0^{(m)}, \O_0^{(k')}, \O_0^{(\a)}$ and $\O_0^{(\b)}$ are of relevance, alongwith $\O_0^{(\L)} = 1 - \O_0^{(m)} - \O_0^{(k')} - \O_0^{(\b)}$ (by Eq.\,(\ref{constraint})).
\eed
Note that the categorization $F_n$ and $C_n$ (where $n = 0,1,2$) imply the spatially flat and curved Universe scenarios respectively. In particular, $F_0$ and $C_0$ correspond to the standard spatially flat and curved $\L$CDM models, for both of which the estimated value of $\O_0^{(m)}$ is around $0.25$ or so, up to the $1\s$ error margins. For all the other scenarios, the corresponding estimates show much reduced best fits of $\O_0^{(m)}$, albeit with about an order of magnitude enhancement in the corresponding $1\s$ error margins. However, even with such enhanced error margins, the reduction in $\O_0^{(m)}$ is quite accountable. To be specific, given that there is about $5\%$ of visible (baryonic) matter in the Universe, the CDM proportion in $F_0$ and $C_0$ is about $20\%$ at the present epoch $z = 0$, as the total (visible + dark) matter density parameter $\O_0^{(m)} \simeq 0.25$, up to $1\s$. In contrast, in the scenarios $F_2$ and $C_2$ for instance, the best fit $\O_0^{(m)}$ values quoted in Table\,\ref{LP-EstTable-2f}, viz. $0.147$ and $0.137$ respectively, imply that the most likely CDM proportions are about $9.7\%$ and $8.7\%$ respectively. Even with the corresponding (positive) $1\s$ errors quoted therein, viz. $0.062$ and $0.57$ respectively, the CDM proportions could be at most $15.9\%$ and $14.4\%$ respectively, still quite lower than $20\%$ for $F_0$ and $C_0$. Thus, we have a clear observational support for a substantial modification of the standard (flat or curved) $\L$CDM model for the varying $G$ of the sort considered here, as a consequence of which we have the terms with the density parameters $\O_0^{(\a)}$, $\O_0^{(\b)}$ and possibly more so, apart from $\O_0^{(\L)}$, which are not kept in the cosmological equation (\ref{hub1b}), for brevity. As such, it is reasonable to go along with a further proposition that it is the logarithmic term in Eq.\,(\ref{hub1b}), with the associated density parameter $\O_0^{(\a)}$, which being present in all the scenarios $F_1$, $F_2$, $C_1$ and $C_2$, compensates for the CDM content of the Universe, and that too by quite an extent, compared to the $\L$CDM cases $F_0$ and $C_0$. On the other hand, the effective dark energy (DE) component of the Universe may be reckoned in this paradigm as
being given by the $\O_0^{(\L)}$ and $\O_0^{(\b)}$ terms in Eq.\,(\ref{hub1b}), 
which are attributed solely to the variation of $G$. 
Considering $\O_0^{(\L)}$ as the principal DE density contributor, one may therefore interpret the `extra mass' or a putative $\L$ of unknown origin, seemingly necessary for explaining a host of phenomena at various scales, 
by a bit higher value of $G$, and consequently stronger gravity at those scales. 
The $\O_0^{(\b)}$ term in Eq.\,(\ref{hub1b}) may, in this context, be considered as that which, if non-zero, makes an additional contribution to the effective DE, and thereby renders the latter {\em dynamical}.  

As to the selection of the favoured scenarios from the above sets $F_n$ and $C_n$, it is imperative to adopt certain criteria, the common ones in the literature being the following:
\bit 
\item Goodness of fit criterion $\,\chi^2_{\text{\scriptsize min}}/{\text{\small dof}}\,$, where $\chi^2_{\text{\scriptsize min}}$ denotes the minimized $\chi^2$ value for the statistical data analysis under consideration, and dof stands for the number of degrees of freedom in such an analysis, i.e. the difference between the number of data points ($N$) in the data-set being used and the number of dimensions or free parameters ($n$) in the model being fitted to the data. Explicitly,
\be \label{chi2dof}
\chi^2_{\text{\scriptsize min}}/{\text{\small dof}} \,:=\, \fr{\text{Minimized} \, \chi^2}{N -\, n} \,.
\ee
\item Akaike information criterion (AIC), which is defined as
\cite{AICref}
\be \label{AIC}
\mbox{AIC} \,:=\, \chi^2_{\text{\scriptsize min}} +\, \fr{2 n N}{N - n - 1} \,.
\ee 
\item Bayesian information criterion (BIC), which is defined as
\cite{BICref}
\be \label{BIC}
\mbox{BIC} \,:=\, \chi^2_{\text{\scriptsize min}} +\, n \, \ln\, N \,.
\ee 
\eit 
The preferred model is the one with the minimal value for each of these criteria. However, it may be the situation that among a given set of models, the one with, say, the minimal $\chi^2_{\text{\scriptsize min}}/{\text{\small dof}}$ or AIC is not the same as that with the minimal BIC. So, one has to be cautious in making a model selection on the basis of these criteria. While there is no general guideline as to which of these criteria should ideally be adopted, the BIC is often considered for the purpose. Nevertheless, there are counter-arguments in that direction as well, since the BIC penalises the models with more free parameters (i.e. larger $n$) very heavily, compared to the others, especially for a data-set of a very large size $N$ (for e.g. the SNeIa Pantheon+SH0ES one, for which $N = 1701$). Moreover, a given model, say $A$, cannot always be completely disfavoured or ruled out, compared to another model, say $B$, even if the AIC or BIC for $A$ exceeds the corresponding value (AIC or BIC) found for $B$. In fact, given such a situation (in which $B$ has a lower AIC or BIC than that for A), both the models $A$ and $B$ may by and large be inferred as being equally favoured by the data, as long as $\, \D AIC = AIC_A - AIC_B \lesssim 2$. Even for a $\D AIC$ up to about $10$, the model $A$ cannot be disfavoured completely in comparison to the model $B$. A value of $\D AIC$ beyond that, of course, would mean ruling out $A$ compared to $B$. Similar rules hold for $\, \D BIC = BIC_A - BIC_B$ as well, see e.g. the section 4 of ref.
\cite{SVP-AICBIC}
or the section 3 of ref.
\cite{FCM-AICBIC}
for detailed discussions on this.

From the $\chi^2_{\text{\scriptsize min}}/{\text{\small dof}}\,$ and the AIC and BIC values quoted in Table\,\ref{LP-EstTable-2f}, for the Pantheon+SH0ES and OHD joint analysis for the various cosmological scenarios emerging from the variation of $G$ considered in this paper, we see that the scenarios $F_2$ and $C_2$ appear to be the most favoured ones in the spatially flat and curved Universe categories respectively. In fact, it is quite remarkable that even with two extra parameters, $\O_0^{(\a)}$ and $\O_0^{(\b)}$, the BIC values of $F_2$ and $C_2$ are lower than the corresponding ones for the spatially flat and curved $\L$CDM scenarios $F_0$ and $C_0$. Among the parameters $\O_0^{(\a)}$ and $\O_0^{(\b)}$ though, it is the latter which is mainly responsible for a drastic reduction of the $\chi^2_{\text{\scriptsize min}}$, and hence of both the AIC and BIC values, in $F_2$ and $C_2$, compared to $F_0$ and $C_0$ respectively. This is evident from the statistics shown in Table\,\ref{LP-EstTable-2f} for the scenarios $F_1$ and $C_1$ wherein the parameter $\O_0^{(\b)}$ not being considered, the presence of the parameter $\O_0^{(\a)}$ alone does not lead to much reduction in $\chi^2_{\text{\scriptsize min}}$. As such, the AIC values for $F_1$ and $C_1$ are within proximity of about $\pm 2$ from those for $F_0$ and $C_0$ respectively, whereas the BIC values for $F_1$ and $C_1$ get increased from those for $F_0$ and $C_0$ by quite some amount ($\sim 7$ and $\sim 5$ respectively). Although not ruled out, $F_1$ and $C_1$ are certainly disfavoured in comparison to the $\L$CDM scenarios $F_0$ and $C_0$, which are themselves not as favoured as $F_2$ and $C_2$ respectively. 

%
\begin{center} 
%
\begin{figure}[p]
\centering
\includegraphics[height=4in,width=4.5in]{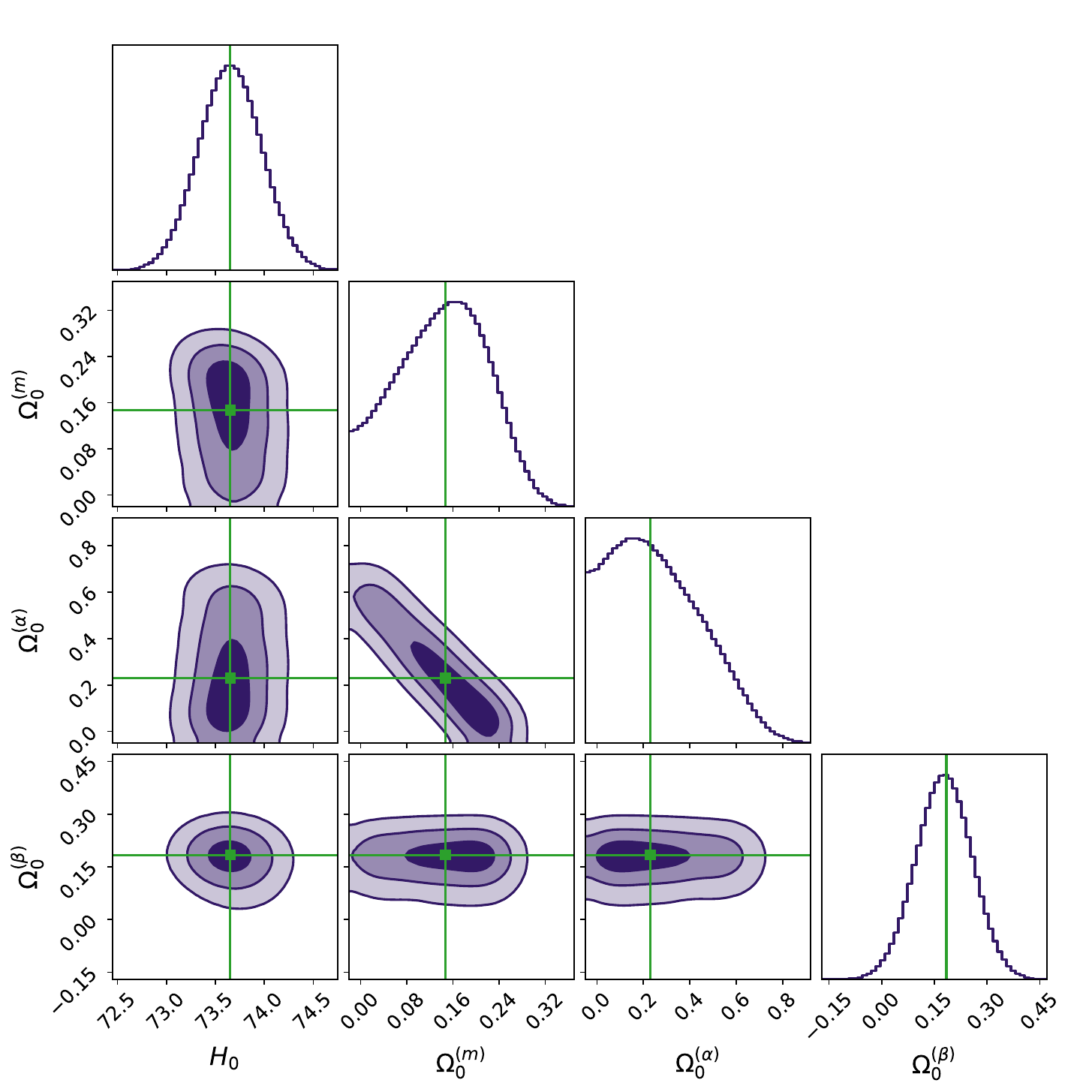}
\caption{\footnotesize The $1\s$-$3\s$ parametric contour levels for the case $F_2$ using the Pantheon+SH0ES data combined with OHD.}
\label{LP_HSN_F2_Fig}
\end{figure}
%
%
%
\begin{figure}[p]
\centering
\includegraphics[height=4.25in,width=4.75in]{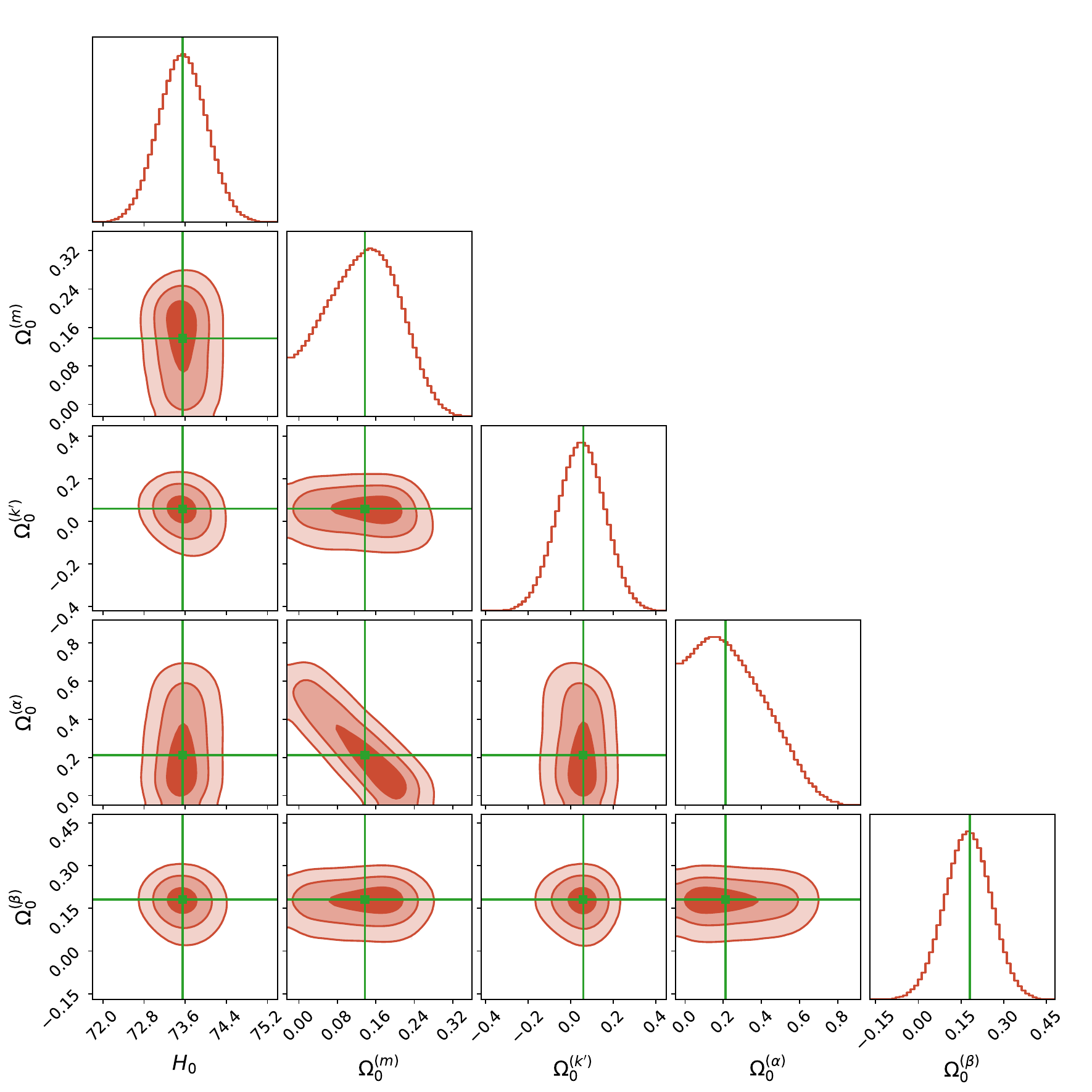}
\caption{\footnotesize The $1\s$-$3\s$ parametric contour levels for the case $C_2$ using the Pantheon+SH0ES data combined with OHD.}
\label{LP_HSN_C2_Fig}
\end{figure}
%
\end{center}
%
%
%
%
%

%
Figs.\,\ref{LP_HSN_F2_Fig} and \ref{LP_HSN_C2_Fig} show the one-dimensional and two-dimensional posterior plots, obtained from the Pantheon + SH0ES and OHD joint analysis for the most favoured scenarios $F_2$ and $C_2$ respectively. While the parametric best fits, in these two scenarios, are marked in the corresponding one-dimensional histogram panels by the vertical lines therein, such markings are done by the dotted intersections of the vertical and horizontal lines in the two-dimensional posterior contours, which are shown up to $3\s$. With $\O_0^{(\a)}$ being conjectured to be compensating for a good chunk of the CDM content of the universe, as the estimates in Table\,\ref{LP-EstTable-2f} and correspondingly the posterior plots in Figs.\,\ref{LP_HSN_F2_Fig} and \ref{LP_HSN_C2_Fig} illustrate, the fairly high values of $\O_0^{(\b)}$ (best fits for both $F_2$ and $C_2$ around $0.18$) imply that the effective DE component is not that weakly dynamical, as is generally assumed (see for e.g. ref.
\cite{SSASB-MST}).
%

%
It also appears from Table\,\ref{LP-EstTable-2f} and Figs.\,\ref{LP_HSN_F2_Fig} and \ref{LP_HSN_C2_Fig} that the well-known Hubble constant ($H_0$) tension problem, that persists for $\L$CDM, is alleviated only by a very slight margin in the scenarios $F_2$ and $C_2$ (see the slightly reduced $H_0$ best fits in these scenarios, compared to that for $F_0$ and $C_0$, quoted in Table\,\ref{LP-EstTable-2f}). Nevertheless, one needs to keep in mind that all the above results pertain to cosmology in a quasi-Newtonian treatment, the primary objective of which is just to have the basic idea as to how things would go at the background cosmological level, for the $G$-variation of the sort considered here. The eventual task is to resort to an appropriate covariant formulation, and analyse the cosmological perturbation spectrum, for a reconciliation of our proposal by examining the imprint of the baryon acoustic oscillations (BAO) on the cosmic microwave background (CMB) and the large-scale structure (LSS). While such an analysis is being carried out in the ongoing work 
\cite{ds2},
let us end this section with the following qualitative remarks from the picture emerged thus far:
\begin{itemize}
\item The evolution of the baryons, and the fluctuations thereof, being grossly affected by an additional gravitational potential, mainly the logarithmic one in our formalism, the acoustic oscillations in the early universe, and their subsequent suppression at smaller scales, would no longer be reckoned in the way as in the standard $\L$CDM model. 
\item Studying the baryon evolution and the acoustic oscillations in an orthodox manner, subject to the $G$-variation, can in fact lead to difficulties. First of all, as pointed out in ref.
\cite{PS},
the baryon transfer function would then be required to account for most of the changes in the baryon perturbations, in the entire red-shift ($z$) range post recombination, if the actual CDM contribution is nominal (as per the above statistical parametric estimates, albeit in the quasi-Newtonian framework and with the low-$z$ data). Secondly, the background dilution of the baryon density and the radiation density not being the standard $\sim (1 + z)^3$ and $\sim (1 + z)^4$ respectively, the matter-radiation equality epoch would get altered, thus modifying the standard expressions for the CMB shift parameter $\cR$, the sound horizon radius $r_s$ at the decoupling red-shift $z_{\text{dec}}$ and at the Compton drag red-shift $z_{\text{drag}}$, and finally the relative BAO distance $r_{\text{\tiny BAO}}$ (for details, see e.g., the Chapter 5 of ref.
\cite{AT-book}).
So, the entire cosmological parametric analysis with the high-$z$ data, for e.g. the BAO data, need to be revoked from the scratch, which is formidable enough in itself.  
\item Nevertheless, it may worth resorting to an effective picture in which the baryons are thought of as being isolated, i.e. having the standard $\sim (1 + z)^3$ background dilution and with the usual acoustic oscillations, while the effect of the logarithmic and the subsequent sub-leading terms in the varying $G$ are supposed to have certain field artifacts, whose fluctuations obviously evolve gravitationally. Some such fluctuations may grow with time, thus suppressing the acoustic signatures in the galaxies, and implying that the corresponding field artifacts differ from, e.g., the scalar field sources of a dynamical dark energy, such as quintessence, whose perturbations oscillate with extremely small amplitudes, thereby getting suppressed compared to the matter perturbations at the sub-horizon scales.
%
%
\end{itemize}

\section{Implications of the varying $G$ at very short distances} \label{sec:VarGshortdist}

Another intriguing aspect of an $r$-dependent Newton's constant is that it can in principle make the law of gravity 
an `inverse $r^k$ law' of the form analogous to a Laurent series:
%
%
%
%
\begin{eqnarray}
E (r) 
%
&=& -\frac{G'M}{r^k} 
\nn \\
&=& 
- \frac{G_{0}'M}{r^k} 
- \frac{G_1' M}{r^{k-1}} + \dots  
%
- \frac{G_{k-2}'M}{r^2} - \dots  ~
%
%
\label{er2} 
\end{eqnarray}
%
where we have put primes ($'$) on the expansion coefficients to distinguish from the expansion (\ref{g1}) and the subsequent steps until now.
The $1/r^2$ term signifies the standard inverse-squared term and $G_{k-2}'$ the standard Newton's constant. In the above, the index $k$ can be very large, with the inverse cubed, inverse fourth power etc. effective at scales much smaller than the currently measured smallest length of about a millimetre, up to which the inverse square law has been tested
\cite{aspel}. 
In this case, the effective (varying) Newton's constant is given by 
%
\be
G'(r) = G_{0}' + G_{1}'r + G_2'r^2 + \dots + G_{k-2}' r^{k-2} + \dots  \,,
\ee
which reveals something remarkable, that the `starting gravitational constant' $G_0'$, effective at short distance scales, can be small, negligible or even zero, building up to the measured value of Newton's constant $G_{k-2}'$ at macroscopic distances. Therefore, when quantum effects are expected to be dominant, i.e. for $r \to 0$, a de facto vanishing coupling constant would preclude any problematic quantum gravity effects, or for that matter any quantum gravity effects\,! If true, this ought to be examined in detail, especially in the context of the various formulations of quantum gravity
\cite{DFS-VarG1,DFS-VarG2}.

\section{Summary and Conclusions} \label{sec:Concl}

Let us first summarize by saying that in this work we have made a fairly intensive study of the implications of a proposed scale-dependence of the Newton's constant $G$, with the objective of finding possible ways of coping with some major challenges posed to the standard Newtonian gravitational theory or Einstein's General Relativity (GR), which expose the limitations of the same. While our proposal simply reckons a radial variation of $G$, in the form of a Taylor expansion about $r = 0$, the terms emerging thereby, in the gravitational potential, do posit as alternatives to widely accepted entities, notably, dark matter (DM) and dark energy (DE), which have originally been conjectured for explaining well-known observed effects, viz. the flatness of galaxy rotation curves and the dimming of Supernovae of type Ia (SNeIa), within the standard (Newtonian or GR) paradigm. In fact, 
such terms can potentially explain a host of observational results 
spanning from the planetary and astrophysical scales, all the way to the cosmological scale. 

We have discussed, in an itemized way in section \ref{sec:VarG-effects}, the arguments or evidences in support of the overall accountability or the degree of robustness of our proposed radial expansion of $G$, from the perspective of astrophysical observations pertaining to the galaxy rotation curves, gravitational lensing and virial theorem, as well as the emerging cosmological picture(s). However, it is worth pointing out that the observations circumventing gravitational lensing by the bullet cluster in particular, have often been claimed to provide a strong evidence in favour of dark matter, as opposed to competing theories, such as MOND. While this may seem to be against our proposal as well, note that a careful examination reveals that important aspects of standard Newtonian/Einstein gravity, such as the Poisson equation, have been assumed all along in the analyses behind such a claim 
\cite{bullet1,bullet2}. 
In any case, although we have mentioned in section \ref{sec:VarG-effects} that the equations (\ref{delta4})-(\ref{epsilon}) apply to all galaxies and clusters, including the bullet cluster, we admit that a more detailed work needs to be done to reconcile our proposal from such a standpoint.

The cosmological implications of the varying $G$, on the other hand, have been examined quite rigorously in section \ref{sec:VarG-cosm}, albeit in a quasi-Newtonian framework. We have worked out the corresponding Friedmann and Raychaudhuri equations (or, rather, the analogues thereof), limiting our attention to the $G$-expansion upto the cubic order in $r(t)$, with the sign of the cubic term already chosen to be such that the resulting correction term in the Friedmann equation resembles a positive cosmological constant. Consequently, the parametric marginalizations have been carried out, using standard statistical techniques and the combination of the widely used SNeIa Pantheon+SH0ES data-set and the observational Hubble data-set, for various effective cosmological scenarios that emerge from the $G$-expansion. Such scenarios include the concordant (spatially flat and curved) $\L$CDM models as well as those with the density parameters defined in terms of the constants $\a$ and $\b$, whose estimates show the extent of the cosmological effects the logarithmic potential term and the constant force term, respectively, arising from the varying $G$. A comparative study of these scenarios is then carried out by resorting to three statistical indicators, viz. the minimized $\chi^2$ per degree of freedom (dof), AIC and BIC. As it turned out, the cosmological scenarios with non-zero $\b$-term (alongside a non-zero $\a$ term), in both the spatially flat and curved universe categories, are the ones most favoured by the observational data under consideration. The parameter estimation showed that the contribution of the cold dark matter (CDM) to the overall energy budget of universe is significantly reduced in both these scenarios, in comparison to the corresponding (spatially flat and curved) $\L$CDM ones. On the other hand, the $\b$-term's contribution 
is by no means nominal, which implies a fairly strong support for an effective {\em dynamical} DE, or its alternative due to the varying $G$. 

Nevertheless, it must be conceded that the quasi-Newtonian formalism, although leads to evolution equations same as those obtained general relativistically, at the background cosmological level, is nothing but heuristic. So there is a stringent need for an embedding of the $G$-variation in a generally covariant theory. A concrete theory of the sort in which such an embedding is fairly straightforward could be, for instance, the non-local gravity (NLG) theory developed in refs.
\cite{HM-NLG,BCHM-NLG,mash-NLG}.
We have briefly illustrated this in the Appendix, particularly from the point of view of the emerging cosmological picture. As inferred therein, a suitable choice of the non-locality measure, in the form of the so-called gravitational susceptibility function $S(t)$, can make the corresponding cosmological equations resemble precisely as the quasi-Newtonian ones. As such, the parametric estimates would be the same as that in section \ref{sec:VarG-cosm}, thus implying a cosmic evolution with a reduced amount of CDM than in the concordant models. However, as mentioned in the ending part of section \ref{sec:VarG-cosm}, a deeper insight to the cosmological background solution as well as the perturbation spectrum is required, in order to account for the CMB and BAO observations. Furthermore, the NLG theory may not be deemed the best one for the covariant embedding of the varying $G$. Therefore, among possible alternatives it is necessary to ascertain which one(s) is(are) the most suitable for the purpose.  

Before concluding, let us also emphasize on the need for careful measurements of planetary scale effects, such as the precession angle of the perihelia of the orbits of planets. In fact, this is the first-up requirement in the sequence of observable effects at progressively higher length scales. Larger the orbit, i.e. more distant the planet from the gravitating source (star), larger would be the expected fractional corrections to the precession amount due to the new terms in the gravitational potential due to the varying $G$. This can be verified by explicit calculations at least in the case in which the (leading) logarithmic correction term in the potential is considered. Specifically, the ratio of the perihelion precession due to the logarithmic potential and that due to the standard general relativistic one, per revolution, turns out to be of the order of $\,a^2/(R_S\,r_1)$, where $a$ is the semi-major axis of the planet and $R_S \approx 3\,$Km is the Schwarzschild radius of the star. This translates to an amount $\sim\,10^{-5}-10^{-3}$ from the nearest to the farthest planets in our solar system, contrary to the most prominence for the nearby planets in the standard general relativistic case. While works are in progress in this direction
\cite{ds2},
we shall not make any further comments here, apart from that the results we have obtained so far fairly corroborate to those in the existing literature (see for e.g.
\cite{LR-PP,DX-PP}
and references therein). The same is the situation for the possible short distance effects of the varying $G$, discussed briefly in section \ref{sec:VarGshortdist}. We have a few works in progress in that direction as well
\cite{DFS-VarG1,DFS-VarG2},
the outcomes of which we hope to report soon.

On the whole, let us conclude by reminding one that whatever presented in this paper is just an initial course or a preliminary attempt to explain observations 
by means of a simple modification of the gravitational field potential. For the effects we consider, viz., rotation curves of galaxies, gravitational lensing, virial theorem and the cosmological background evolution, the emerging picture does seem to hold up well. As to the related subtle issues, as well as the stringent ones, if indeed all of these come through, then that would signal the cure for major gravitational maladies we know of, at scales ranging from the planetary to the cosmological, perhaps without requiring to go with the DM or(and) DE conjecture(s) at all! Nevertheless, for a complete understanding of the
implications of our proposal, it again comes back to the question of embedding the varying $G$ within a covariant framework and to assert which such framework(s) is(are) the most suitable for the same. As already mentioned, this is being examined in the ongoing works
\cite{ds2,SAD}
to be reported upon completion.  
%
%
%
%
%

%
%
%
%


\section*{Acknowledgment}

This work was supported by the Natural Sciences and Engineering Research Council of Canada. SS acknowledges financial support from the Faculty Research Programme Grant -- IoE, University of Delhi (Ref.No./ IoE/ 2024-25/12/FRP). 
%
The authors are grateful to the anonymous reviewer of the paper, for making several important suggestions and bringing in notice a wealth of valued references.


\section*{Data Availability Statement}
The manuscript has no associated data or the data will not be deposited. Observational data-sets used for the statistical estimation are retrieved from well-known references, duly cited in this paper.


\section*{Appendix: Covariant embedding of the varying $G$ --- an illustration}

Following refs.
\cite{HM-NLG,BCHM-NLG,mash-NLG},
consider the non-local extension of the teleparallel equivalent of General Relativity (TEGR), which we shall refer henceforth as the non-local gravity (NLG) theory. The formalism is essentially developed in analogy with the non-local electrodynamics of media, wherein the constitutive relations between the electromagnetic excitation $H^{\a\b} = (\vD,\vH)$ and the field strength $F_{\a\b} = (\vE,\vB)$ depend on the evolution history and are hence non-local:
\be \label{NL-EM}
H^{\a\b} =\, \sq{-g} \,g^{\a\c} g^{\b\d} F_{\c\d} (x) +\, \int d^4 x' \chi^{\a\b\c\d} (x,x')\, F_{\c\d} (x') \,,
\ee 
where $g$ denotes the determinant of the metric tensor $g_{\a\b}$ and $\chi^{\a\b\c\d}$ is a kernel which determines the degree of non-locality of the electromagnetic configuration. 

Teleparallelism, on the other hand, concerns 
preferred orthonormal tetrad frames, such that
\be \label{tetg}
g_{\m\n} (x) =\, \eta_{\ha\hb} \, e_\m^{~\ha}(x) \, e_\n^{~\hb}(x) \,,
\ee
where $e_\m^{~\ha}$ denotes the tetrad field, and $\eta_{\ha\hb}$ is the Minkowski metric of the local tangent space, with $\ha,\hb,\dots$ being the corresponding coordinate labels, different from $\a,\b,\m,\n,\dots$, which are the usual curved space coordinate labels. Defining an asymmetric affine connection, known as the {\em Weitzenb\"ock connection}
\be \label{WC}
\C^\s_{\m\n} =\, e^\s_{~\ha} \, \pa_\m e_\n^{~\hb} \,,
\ee
one perceives curvature-freeness, i.e. the space-time is parallelizable, in accord with $\nabla_\n e_\m^{~\ha} = 0$, and as such, the metricity condition $\nabla_\n g_{\a\b} = 0$, where $\nabla_\n$ denotes the covariant derivative in terms of $\C^\s_{\m\n}$ given by eq.\,(\ref{WC}). The space-time nonetheless admits a non-zero torsion
\be \label{W-tors}
C_{\m\n}^{~~\s} := \, 2 \,\C^\s_{[\m\n]} =\, 2 \,e^\s_{~\ha} \,\pa_{[\m} \,e_{\n]}^{~\ha} \,,
\ee 
which is nothing but the gravitational field strength, similar to the field strength $F_{\a\b} = 2 \pa_{[\a} A_{\b]}$ in electrodynamics (with $A_\b$ being the corresponding field potential). On the whole, the teleparallel gravitational formulation is that of a translational gauge theory --- an Abelian one, analogous to electrodynamics. Such a formulation is in fact equivalent to General Relativity (GR) in the sense that it leads to the same field equations, with the gravitational gauge field strength given by the Einstein tensor expressed in terms of the tetrads and the Weitzenb\"ock connection as
\bea \label{Einst}
G_{\m\n} &=& \fr{\k^2}{\sq{-g}} \bigg[e_\m^{~\hc} \,g_{\n\a} \,\pa_\b \cH^{\a\b}_{~~~~\hc} \nn\\
&& - \le(C_\m^{~\a\b} \cH_{\n\a\b} - \fr 1 4 C^{\a\b\s} \cH_{\a\b\s}\ri)\bigg] \,,
\eea 
where $\k^2 = 8 \pi G$ (with $G$ being the usual Newton's constant), and 
\bea 
&&\cH_{\m\n\s} =\, \fr{\sq{-g}}{\k^2} \cC_{\m\n\s} \,,
\label{constrel} \\
&&\cC_{\m\n\s} =\, 2 C_{[\m} \,g_{\n]\s} -\, C_{\m[\n\s]} - \fr 1 2 C_{\n\s\m} \,, 
\label{modtor} \\
&&C_\m \equiv\, C^\a_{~\m\a}~:~\text{torsion trace vector.}
\eea 
In analogy with the electrodynamics of media, Eq.\,(\ref{constrel}) is generally considered as a constitutive relation, with $\cH_{\m\n\s}$ treated as the gravitational excitation from the gauge theoretic perspective. The non-local generalization to this, proposed in a series of papers
\cite{HM-NLG,BCHM-NLG,mash-NLG},
is an augmentation of the modified torsion tensor $\cC_{\m\n\s}$ on the right hand side of Eq.\,(\ref{constrel}) with a non-local term $N_{\m\n\s}$ whose tangent space projection is expressed as
\be \label{NLterm}
N_{\hmu\hn\hs} =\, \int d^4 x' \sq{-g(x')} \, \chi(x,x') \, X_{\hmu\hn\hs} (x') \,,
\ee 
with $\chi$ being a causal kernel, and
\be
X_{\hmu\hn\hs} =\, \cC_{\hmu\hn\hs} +\, q \, \Ct_{[\hmu} \,\eta_{\hn]\s} \,,
\ee 
where $\Ct_{\hmu}$ denotes the tangent space projection of the torsion pseudo-trace vector
\be 
\Ct_{\m} \equiv\, \fr 1 3 \, \e_{\a\b\c\m} C^{\a\b\c} \,,
\ee
and $q$ is a dimensionless constant determining the extent of the gravitational parity violation.

The Newtonian limit of the NLG theory leads to a modified Poisson's equation which corresponds precisely to a logarithmic correction term in the gravitational potential we are dealing here in the context of the $G$-variation. The NLG Newtonian regime has therefore been of interest and extensive studies have been carried out from the point of view of explaining the flat rotation curves of galaxies
\cite{RM-NLG,BM-NLG,RoM-NLG}.
However, beyond linearization, the (fully covariant) NLG theory has no known non-linear exact solution till date. As such, the focus has shifted in recent times to the study of the local limit of the non-local TEGR, construed from the following form of the NLG kernel
\cite{TBM-NLG,TBBM-NLG-1,TBBM-NLG-2}:
\be \label{NLG-kern}
\chi (x,x') =\, \fr{S(x)}{\sq{-g(x)}} \d(x - x') \,,
\ee
where $S(x)$ is a dimensionless scalar function of coordinates, which is referred to as the gravitational {\em susceptibility function}, in analogy with electrodynamics. 

For the spatially homogeneous cosmological space-times, the susceptibility function can be taken to be varying with time only.  The non-local TEGR equations lead to the modified Friedmann equations:
\bea 
&& (1 + S) \le(\fr{{\dot a}^2}{a^2} +\, \fr k {a^2}\ri) =\, \fr{8 \pi G} 3 \, \r \,,
\label{NLG-Feq1} \\
&& (1 + S) \le(\fr{2 \ddot a} a +\, \fr{{\dot a}^2}{a^2} +\, \fr k {a^2}\ri) =\, - 8 \pi G p -\, 2 \dot{S} \fr{\dot{a}} a \,, \qquad
\label{NLG-Feq2}
\eea
where $\r$ and $p$ denote, respectively, the total energy density and the total pressure, and $k = 0, \pm 1$ is the spatial curvature constant. In addition, there are two constraints, viz.
\be \label{NLG-constraints}
S > -1 \,, \quad k \dot{S} = 0 \,.
\ee 
Therefore either (i) $\, k = \pm 1$ and $\,S =$ constant, or (ii) $k = 0$ and $S$ is an arbitrary function of time. While the case (i) implies the same closed and open Friedmann solutions, modulo a numerical scaling of $\r$ and $p$ by a factor $(1 + S)^{-1}$, the case (ii) is potentially of much interest as it not only complies with the assumption of the spatial flatness of the universe, which the observations grossly indicate, but also marks the outcome of an effective $G$-variation, since the above equations (\ref{NLG-Feq1}) and (\ref{NLG-Feq2}) can be recast as the following set:
\bea 
&& H^2 =\, \fr{8 \pi \Geff} 3 \, \r \,,
\label{NLG-Feq1a} \\
&& \dot H =\, - 4 \pi \Geff \le(\r + p\ri) -\, \fr{\dot{G}_{\text{eff}}}{\Geff} \, H  \,, \qquad
\label{NLG-Feq2a}
\eea
where 
\be 
\Geff(t) =\, \fr G {1 + S(t)} \,.
\ee 
As such, choosing suitably the function $S(t)$, and hence $\Geff(t)$, one can reproduce the cosmological equations (\ref{hub1}) and (\ref{R-eq}) obtained in the quasi-Newtonian framework (upon replacing $G$ by $G_0$ and $\Geff(t)$ by $G(t)$). Nevertheless, whether such a choice is justifiable from a physical standpoint is presently being examined, among several issues, in the ongoing works
\cite{ds2,SAD}.
%



\begin{thebibliography}{99}

\bibitem{mikkelsen}
D.R. Mikkelsen and M.J. Newman, 
Phys. Rev. {\bf D 16} (1977) 919. 

\bibitem{sivaram}
C. Sivaram, K. Arun and L. Rebecca, 
J. Astrophys. Astron. {\bf 41} (2020) 1, 4.

\bibitem{bd1}
C. Brans, R.H. Dicke, 
Phys. Rev. {\bf 124} (1961) 925.

\bibitem{bd2}
R.H. Dicke, 
Phys. Rev. {\bf 125} (1962) 2163 .

\bibitem{FM-ST} 
Y. Fujii and K. Maeda, {\em The Scalar-Tensor Theory of Gravitation}, Cambridge Monographs on Mathematical Physics, Cambridge University Press, United Kingdom (2003).

\bibitem{frni-ST}
V. Faraoni, {\em Cosmology in Scalar-Tensor Gravity}, Kluwer Academic Publishers (2004).

\bibitem{fr}
T.P. Sotiriou and V. Faraoni, Rev. Mod. Phys. {\bf 82} (2010) 451.

\bibitem{CFPS-mg}
T. Clifton, P. G. Ferreira, A. Padilla and C. Skordis, Phys. Rept. {\bf 513} (2012) 1.

\bibitem{papa-ed}
E. Papantonopoulos, Lecture Notes in Physics, Springer, Switzerland (2015).

\bibitem{NOO-mg}
S. Nojiri, S.D. Odintsov and V.K. Oikonomou, Phys. Rept. {\bf 692} 
(2017) 1.

\bibitem{bek-TeVeS}
J.D. Bekenstein, Phys. Rev. {\bf D 70} (2004) 8, 083509.

\bibitem{MSY-TeVeS}
N.E. Mavromatos, M. Sakellariadou and M.F. Yusaf, Phys. Rev. {\bf D 79} (2009) 8, 081301.

\bibitem{BS-TeVeS}
J.D. Bekenstein and R.H. Sanders, Mon. Not. Roy. Astron. Soc. {\bf 421} (2012) L59.

\bibitem{mil-MOND}
M. Milgrom, Astrophys. J. {\bf 270} (1983) 365.

\bibitem{MS-MOND}
M. Milgrom and R.H. Sanders, Astrophys J. {\bf 599} (2003) 1, 25.

\bibitem{bek-MOND}
J.D. Bekenstein,  Phys. Rev. {\bf D 70} (2004) 83509.

\bibitem{HM-NLG}
F.W. Hehl and B. Mashhoon, Phys. Rev. {\bf D 79} (2009) 064028.

\bibitem{BCHM-NLG}
H.-J. Blome, C. Chicone, F.W. Hehl and B. Mashhoon, Phys. Rev. {\bf D 81} (2010) 065020.

\bibitem{mash-NLG}
B. Mashhoon, Symmetry {\bf 14} (2022) 10, 2116.

\bibitem{nash}
G. Nash, Int. J. Mod. Phys. {\bf D 32} (2023) 06, 2350031.

\bibitem{RW-QGRG}
M. Reuter and H. Weyer, 
Int. J. Mod. Phys. {\bf D 15} (2006) 2011. 

\bibitem{SSS-QGRG}
I.L. Shapiro, J. Sol\`a and H. \v{S}tefan\v{c}i\'c, 
JCAP {\bf 0501} (2005) 012. 

\bibitem{GNS1-QGRG}
B.L. Giacchini, T. de Paula Netto and I.L. Shapiro, 
JHEP {\bf 10} (2020) 011.

\bibitem{GNS2-QGRG}
B.L. Giacchini, T. de Paula Netto and I.L. Shapiro, 
Phys. Rev. {\bf D 102} (2020) 106006. 

\bibitem{RLS-QGRG}
D.C. Rodrigues, P.S. Letelier and I.L. Shapiro, 
JCAP {\bf 1004} (2010) 020. 

\bibitem{DM-QGGF}
D.A.R. Dalvit and F.D. Mazzitelli,
Phys. Rev. {\bf D 56} (1997) 7779.

\bibitem{NMS-QGGF}
T. de Paula Netto, L. Modesto and I.L. Shapiro,
Eur. Phys. J. {\bf C 82} (2022) 2, 160.

\bibitem{BHMR-RGcosm}
N.R. Bertini, W.S. Hip\'olito-Ricaldi, F. de Melo-Santos and D.C. Rodrigues, 
Eur. Phys. J. {\bf C 80} (2020) 479. 

\bibitem{SS-RGcosm}
J. Sol\`a and H. \v{S}tefan\v{c}i\'c,
Phys. Lett. {\bf B 624} (2005) 147.

\bibitem{AKLR-RGcosm}
P.D. Alvarez, B. Koch, C. Laporte and \'A. Rinc\'on,
JCAP {\bf 06} (2021) 019.

\bibitem{BRS-RGcosm}
N.R. Bertini, D.C. Rodrigues and I.L. Shapiro,
Phys. Dark Univ. {\bf 45} (2024) 101502.

\bibitem{dsphysopen}
S. Das and S. Sur, 
Phys. Open {\bf 15} (2023) 100150.

\bibitem{bullet1}
D. Clowe, A. Gonzalez and M. Markevitch, 
Astrophys. J. {\bf 604} (2004) 596.

\bibitem{bullet2}
D. Clowe, A. Gonzalez and M. Markevitch, 
Astrophys. J. {\bf 648} (2004) L109.

\bibitem{binney}
J. Binney, S. Tremaine, 
{\em Galactic Dynamics}, Second Edition, Princeton (2008).

\bibitem{yahalom}
A. Yahalom, 
Symmetry {\bf 12} (2020) 1693. 

\bibitem{inverno}
R. D'Inverno,
{\it Introducing Einstein's Relativity}, Clarendon Press,
Oxford (1992), Chapter 22, pp.\,310-312.

\bibitem{liddle}
A. Liddle, 
{\it An Introduction to Modern Cosmology}, Second Edition,
Wiley (2003), Chapter 3, pp.\,17-24.

\bibitem{rai}
A. Rai Choudhuri, 
{\it Astrophysics for physicists}, Cambridge (2010), 
Chapter 10, pp.\,309-313.

\bibitem{weinberg}
S. Weinberg, 
{\it Cosmology}, Oxford (2008).

\bibitem{ds2}
S. Das and S. Sur, in preparation.

\bibitem{SAD}
S. Sur, R. Aggrawal and S. Das, in preparation.

\bibitem{ds3}
S. Das and S. Sur, 
Phys. Dark Univ. {\bf 42} (2023) 101331.

\bibitem{scol-Pan+} 
D. Scolnic et. al., 
Astrophys. J. {\bf 938} (2022) 113. 

\bibitem{brout-Pan+}
D. Brout et. al., 
Astrophys. J. {\bf 938} (2022) 110. 

\bibitem{riess-SH0ES}
A.G. Riess et. al., 
Astrophys. J. Lett. {\bf 934} (2022) 1, L7. 

\bibitem{Pan+} 
The Pantheon+SH0ES data release plugin with cosmosys chains and likelihoods, is available at https://github.com/PantheonPlusSH0ES.

\bibitem{mors-OHD}
M. Moresco {\it et. al.}, 
JCAP {\bf 08} (2012) 006.

\bibitem{MPCJM-OHD}
M. Moresco, L. Pozzetti, A. Cimatti, R. Jimenez and C. Maraston,
JCAP {\bf 05} (2016) 014.

\bibitem{BMC-OHD}
N. Borghi, M. Moresco and Cimatti, 
Astrophys. J. Lett. {\bf 928} (2022) L4. 

\bibitem{tomasetti-OHD}
E. Tomasetti {\it et. al.}, 
Astron. \& Astrophys. {\bf 679} (2023) A96.




\bibitem{riess-Lad_OHD}
A.G. Riess {\it et. al.},
Astrophys. J. {\bf 853} (2018) 126.

\bibitem{GCH-BAO-Gal_OHD}
E. Gazta\~naga, A. Cabr\`e and L. Hui, 
Mon. Not. Roy. Astron. Soc. {\bf 399} (2009) 1663.

\bibitem{blake-BAO-Gal_OHD}
C. Blake {\it et. al.}, 
Mon. Not. Roy. Astron. Soc. {\bf 425} (2012) 405.

\bibitem{zhao-BAO-QSO_OHD}
G.-B. Zhao {\it et. al.}, 
Mon. Not. Roy. Astron. Soc. {\bf 482} (2019) 3497.

\bibitem{font-BAO-Lya_OHD}
A. Font-Ribera {\it et. al.}, 
JCAP {2014} (2014) 027.

\bibitem{delubac-BAO-Lya_OHD}
T. Delubac {\it et. al.}, 
Astron. \& Astrophys. {\bf 574} (2015) A59.

\bibitem{dong-BAO-Lya_OHD}
C. Dong {\it et. al.}, 
Mon. Not. Roy. Astron. Soc. {\bf 514} (2022) 5493.

\bibitem{monjo}
R. Monjo, 
Astrophys. J. {\bf 967} (2024) 66.

\bibitem{AICref}
H. Akaike, 
IEEE Trans. Autom. Control {\bf 19} (1974) 6, 716.

\bibitem{BICref}
A.R. Liddle, 
Mon. Not. R. Astron. Soc. {\bf 377} (2007) L74.

\bibitem{SVP-AICBIC}
J. Sol\`a, A. G/'omez-Valent and J.r de Cruz P\'erez,
Astrophys. J. {\bf 836} (2017) 1, 43.

\bibitem{FCM-AICBIC}
F. Arevalo, A. Cid and J. Moya, 
Eur. Phys. J. C {\bf 77} (2017) 565.

\bibitem{SSASB-MST}
S. Sur and A. S. Bhatia, 
JCAP {\bf 1707} (2017) 039.



\bibitem{PS}
K. Pardo and D.N. Spergel, 
Phys. Rev. Lett. {\bf 125} (2020) 21, 211101.

\bibitem{AT-book}
L. Amendola and S. Tsujikawa,
{\em Dark Energy: Theory and Observations}, Cambridge University Press, United Kingdom (2010).

\bibitem{aspel}
T. Westphal, H. Hepach, J. Pfaff and M. Aspelmeyer, 
Nature {\bf 591} (2021) 225.

\bibitem{DFS-VarG1}
S. Das, M. Fridman and S. Sur,
in preparation.

\bibitem{DFS-VarG2}
S. Das, M. Fridman and S. Sur,
in preparation.

\bibitem{LR-PP}
L. Iorio and M.L. Ruggiero,
Schol. Res. Exch. {\bf 2008} (2008) 968393.

\bibitem{DX-PP}
X.-M. Deng and Y. Xie,
Ann. Phys. {\bf 361} (2015) 62.

\bibitem{RM-NLG}
S. Rahvar and B. Mashhoon, Phys. Rev. {\bf D 89} (2014) 104011.

\bibitem{BM-NLG}
D. Bini and B. Mashhoon, Phys. Rev. {\bf D 90} (2014) 2, 024030.

\bibitem{RoM-NLG}
M. Roshan and B. Mashhoon, Astrophys. J. {]bf 934} (2022) 1, 9.

\bibitem{TBM-NLG}
J. Tabatabaei, S. Baghram and B. Mashhoon, Mon. Not. Roy. Astron. Soc. {\bf 530} (2024) 795.

\bibitem{TBBM-NLG-1}
J. Tabatabaei, A. Banihashemi, S. Baghram and B. Mashhoon, Int. J. Mod. Phys. {\bf D 32} (2023) 14, 2342009.

\bibitem{TBBM-NLG-2}
J. Tabatabaei, A. Banihashemi, S. Baghram and B. Mashhoon, Astrophys. J. {\bf 965} (2024) 2, 116.



















\end{thebibliography}
\end{document}